\documentclass[manuscript]{aastex63}

\received{***}
\revised{***}
\accepted{***}
\submitjournal{ApJ}
\shorttitle{Triggering process of an X-class flare on small AR 12887}

\shortauthors{Song et al.}

\begin{document}

\title{The triggering process of an X-class solar flare on a small quadrupolar active region}

\correspondingauthor{Jing-Song Wang, Xiaoxin Zhang, Hechao Chen}
\email{wangjs@cma.gov.cn, xxzhang@cma.gov.cn, hechao.chen@pku.edu.cn}

\author[0000-0003-3568-445X]{Qiao Song}
\affiliation{Key Laboratory of Space Weather, National Satellite Meteorological Center (National Center for Space Weather), China Meteorological Administration, Beijing 100081, People’s Republic of China}
\affiliation{Yunnan Key Laboratory of the Solar physics and Space Science, Kunming 650216, People’s Republic of China}
\affiliation{Innovation Center for FengYun Meteorological Satellite (FYSIC), Beijing 100081, People’s Republic of China}

\author{Jing-Song Wang}
\author{Xiaoxin Zhang}
\affiliation{Key Laboratory of Space Weather, National Satellite Meteorological Center (National Center for Space Weather), China Meteorological Administration, Beijing 100081, People’s Republic of China}
\affiliation{Innovation Center for FengYun Meteorological Satellite (FYSIC), Beijing 100081, People’s Republic of China}

\author[0000-0001-7866-4358]{Hechao Chen}
\affiliation{School of Physics and Astronomy, Yunnan University, Kunming 650500, People’s Republic of China} 
\affiliation{Yunnan Key Laboratory of the Solar physics and Space Science, Kunming 650216, People’s Republic of China}

\author[0000-0002-6565-3251]{Shuhong Yang}
\affiliation{National Astronomical Observatories, Chinese Academy of Sciences, Beijing 100101, People’s Republic of China}
\affiliation{School of Astronomy and Space Science, University of Chinese Academy of Sciences, Beijing 100049, People’s Republic of China} 

\author[0000-0003-4804-5673]{Zhenyong Hou}
\affiliation{School of Earth and Space Sciences, Peking University, Beijing 100871, People’s Republic of China}

\author[0000-0002-9534-1638]{Yijun Hou}
\affiliation{National Astronomical Observatories, Chinese Academy of Sciences, Beijing 100101, People’s Republic of China}
\affiliation{School of Astronomy and Space Science, University of Chinese Academy of Sciences, Beijing 100049, People’s Republic of China} 

\author[0000-0003-3605-6244]{Qian Ye}
\affiliation{Key Laboratory of Space Weather, National Satellite Meteorological Center (National Center for Space Weather), China Meteorological Administration, Beijing 100081, People’s Republic of China}
\affiliation{Innovation Center for FengYun Meteorological Satellite (FYSIC), Beijing 100081, People’s Republic of China}

\author{Peng Zhang}
\author{Xiuqing Hu}
\affiliation{Key Laboratory of Radiometric Calibration and Validation for Environmental Satellites, National Satellite Meteorological Center (National Center for Space Weather), China Meteorological Administration, Beijing 100081, People’s Republic of China}
\affiliation{Innovation Center for FengYun Meteorological Satellite (FYSIC), Beijing 100081, People’s Republic of China}
\author{Jinping Dun}
\author{Weiguo Zong}
\affiliation{Key Laboratory of Space Weather, National Satellite Meteorological Center (National Center for Space Weather), China Meteorological Administration, Beijing 100081, People’s Republic of China}
\affiliation{Innovation Center for FengYun Meteorological Satellite (FYSIC), Beijing 100081, People’s Republic of China}

\author[0000-0003-2686-9153]{Xianyong Bai}
\affiliation{National Astronomical Observatories, Chinese Academy of Sciences, Beijing 100101, People’s Republic of China}

\author{Bo Chen}
\author{Lingping He}
\author{Kefei Song}
\affiliation{Changchun Institute of Optics, Fine Mechanics and Physics, Chinese Academy of Sciences, Changchun 130033, People’s Republic of China} 
\affiliation{State Key Laboratory of Applied Optics, Chinese Academy of Sciences, Changchun 130033, People’s Republic of China}

\begin{abstract}
The occurrence of X-class solar flares and their potential impact on the space weather often receive great attention than other flares. But predicting when and where an X-class flare will occur is still a challenge. With the multi-wavelength observation from the Solar Dynamics Observatory and FengYun-3E satellite, we investigate the triggering of a GOES X1.0 flare occurring in the NOAA active region (AR) 12887. Our results show that this unique X-class flare is bred in a relatively small but complex quadrupolar AR. Before the X-class flare, two filaments (F1 and F2) exist below a null-point topology of the quadrupolar AR. Magnetic field extrapolation and observation reveal that F1 and F2 correspond to two magnetic flux ropes with the same chirality and their adjacent feet rooted at nonconjugated opposite polarities, respectively. Interestingly, these two polarities collide rapidly, accompanied by photospheric magnetic flux emergence, cancellation and shear motion in the AR center. Above this site, F1 and F2 subsequently intersect and merge to a longer filament (F3) via a tether-cutting-like reconnection process. As a result, the F3 rises and erupts, involving the large-scale arcades overlying filament and the quadrupolar magnetic field above the AR, and eventually leads to the eruption of the X-class flare with a quasi-X-shaped flare ribbon and a coronal mass ejection. It suggests that the rapid collision of nonconjugated opposite polarities provides a key condition for the triggering of this X-class flare, and also provides a featured case for flare trigger mechanism and space weather forecasting.
\end{abstract}

\keywords{Solar flares (1496); Solar filaments (1495); Solar magnetic fields (1503); Solar active regions (1974); Space weather (2037)}

\section{Introduction} \label{sec:intro}
Solar flares are one of the most explosive phenomena in the Sun's atmosphere and may impact the space weather \citep{2021LRSP....18....4S}. Therefore, the triggering mechanism of major flares ($\geqslant$ M1.0) has been the focus of solar physics researches and the space weather forecasting. Its enormous energy of up to 10$^{32}$ erg comes from the solar magnetic field. The previous studies \citep{2000ApJ...540..583S, 2019JASTP.188...44T} have shown that major flares usually tend to erupt from large active regions (ARs), where the total unsigned magnetic flux is generally large too. Other parameters of the magnetic field, such as the magnetic shear and the magnetic free energy of ARs, are also important tools to analyze major flares \citep{1996ApJ...456..861W, 1999A&A...351..707T, 2007ApJ...656.1173L}. 

The magnetic field continuously emerges from the convective zone to the photosphere of the Sun, and emerging flux regions (EFRs) carry a considerable amount of magnetic flux and magnetic shear, which may effectively enhance the non-potentiality and the occurrence of major flares \citep{2013RAA....13..226S}. Recent research has revealed that collisional shearing of opposite polarities from different EFRs may lead to major flares \citep{2019ApJ...871...67C}. The magnetic flux cancellation \citep{2010ApJ...708..314A, 2011A&A...526A...2G}  and fast rotating sunspots \citep{2007ApJ...662L..35Z} are closely related to the eruptions of flares and coronal mass ejections (CMEs). Sometimes, the various magnetic field characteristics mentioned above are concentrated around the polarity inversion line (PIL) of a $\delta$ sunspot, where the umbrae of opposite polarities in a single penumbra, resulting in the great activity of ARs \citep{1987SoPh..113..267Z,2005ApJ...622..722L,2017ApJ...849L..21Y,2017ApJ...834...56T,2021A&A...648A.106L}. 

Previous studies have shown that the tether-cutting reconnection, kink instability, and torus instability may be responsible for the onset of solar magnetic explosions \citep{2001ApJ...552..833M, 2014ApJ...797L..15C,2018ApJ...869...78C,2019ApJ...887..118C}. The interaction of two filaments supported by two magnetic flux ropes (MFRs) can also lead to an eruption \citep{2001ApJ...553..905L, 2017ApJ...838..131Y}, and some of the observed cases were the interaction or even merging of two filaments with the same chirality \citep{2014ApJ...793...14J, 2017ApJ...836..160Z}, and some were with opposite chiralities \citep{2016ApJ...825..123J}.

Generally, the CSHKP standard flare model \citep{1964NASSP..50..451C, 1966Natur.211..695S, 1974SoPh...34..323H, 1976SoPh...50...85K} is used to explain the magnetic reconnection and the flare process in a two-dimensional (2D) scenario. In the CSHKP model, the core magnetic structure of the eruption (for example, MFRs) located near the PIL rises for some reasons, resulting in a vertical current sheet and the magnetic reconnection below it in the corona \citep{1991ApJ...373..294F}. Particles accelerated by the energy released by the magnetic reconnection will then move downward along the reconnecting magnetic loops to the chromosphere and even to the photosphere, where two gradually separated flare ribbons are formed at the foot points of the magnetic field lines. As a result, these flares are called two-ribbon flares \citep{1997ApJ...474L..61Y,2011SSRv..159...19F}. 

However, recent studies on X-shaped flares highlighted the limitations of the CSHKP model as a 2D model and explored the three-dimensional (3D) nature of the flare process \citep{2016NatSR...634021L, 2017ApJ...850....8J, 2017ApJ...842..106K}. As the term suggests, X-shaped flares display a relatively rare, complex flare ribbon that resembles the letter `X'. The source region for an X-shaped flare is usually a 3D quadrupolar magnetic field, which consists of two oppositely directed dipoles that are not in the same line. It forms magnetic topologies like null points and quasi-separatrix layers (QSLs), where the currents are accumulated and finally lead to magnetic reconnections \citep{2021RSPSA.47700949L}. Quadrupolar fields of different scales may exist at various locations on the Sun. Some quadrupolar magnetic fields were composed of two ARs that were respectively located in the northern and southern hemispheres of the Sun \citep{2014ApJ...787L..27S}, and someones were composed of large-scale magnetic structures among multiple ARs in the same hemisphere \citep{2020A&A...640A.101H}. 
Occasionally, the quadrupolar magnetic configuration was located in a limited part of an AR \citep{2012ApJ...757..149S,2016ApJ...823L..13L, 2022ApJ...926..143M} or at the outskirts of an AR \citep{2016NatSR...634021L}. The quadrupolar magnetic field may constitute the main part of an AR, but it has not been reported yet.

This paper reports a small AR (NOAA AR 12887) mainly constituted by a 3D quadrupolar magnetic field, in which the second X-class flare of the current Solar Cycle 25 erupted. The observations and data related to the X-class flare will be briefed in Section \ref{sec:data}. Section \ref{sec:results} focuses on an interesting question: how a powerful flare was triggered in the small AR? The discussion and conclusions are given in Section \ref{sec:dc}.

\begin{figure*}[!htb]
\plotone{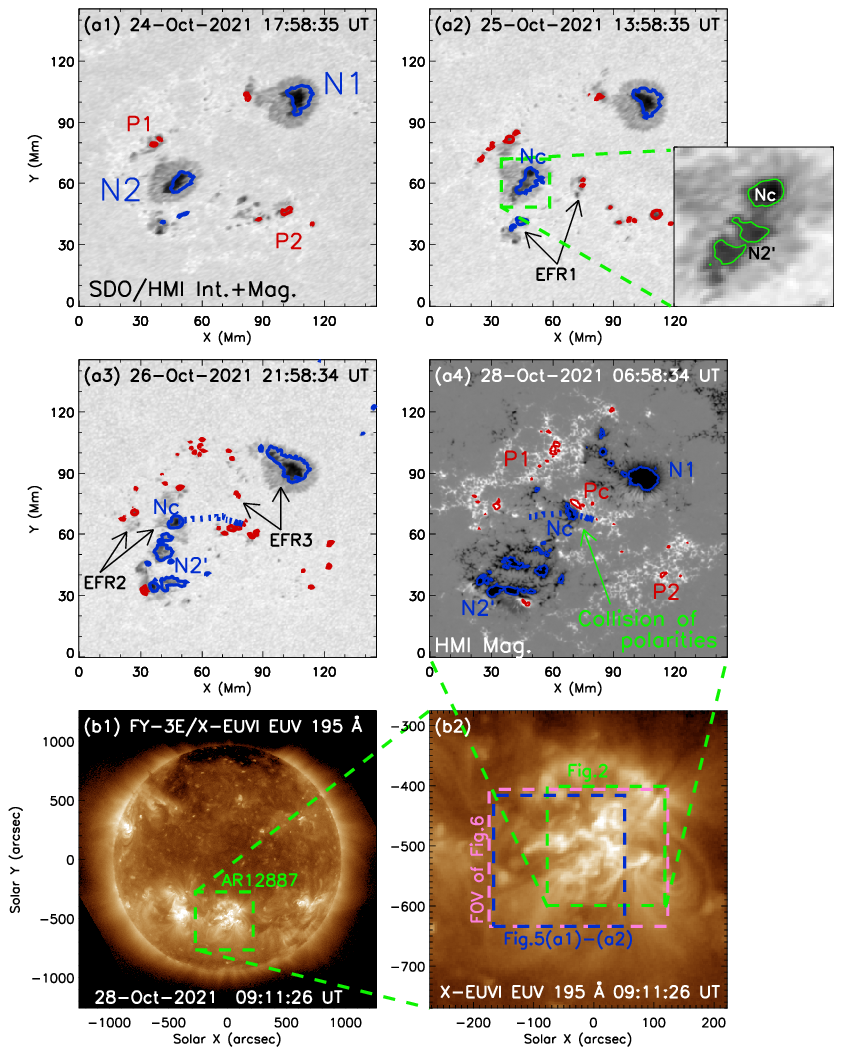}
\caption{Evolution of sunspots and magnetic fields in AR 12887. (a1)-(a3) display SDO/HMI intensitygrams with the contours of the photospheric magnetic field representing the positive (red) and negative (blue) polarities at 800 and $-$800 G, respectively. (a4) specifically gives the magnetic field in the AR on the day of the X1.0 flare eruption, where the meaning of the contours is the same as (a1). (b1)-(b2) show that the AR on October 28, 2021 was located near the center of the solar disk in FY-3E/X-EUVI EUV 195 {\AA} image, and the green box in (b2) gives the field of view (FOV) of Figure \ref{fig:fig2}, while blue and pink boxes show the FOVs of Figure \ref{fig:fig5}(a1)-(a2) and Figure \ref{fig:fig6}, respectively.
\label{fig:fig1}}
\end{figure*}

\section{Observations and data analysis} \label{sec:data}
This study used the ultraviolet and extreme ultraviolet (UV/EUV) multi-wavelength images of Atmospheric Imaging Assembly (AIA, \citealt{2012SoPh..275...17L}) onboard the Solar Dynamics Observatory (SDO, \citealt{2012SoPh..275....3P}), with a cadence of 12 seconds and a spatial resolution of 0.6{\arcsec} pixel$^{-1}$, as well as 195 {\AA} images of the Solar X-ray and Extreme Ultraviolet Imager (X-EUVI, \citealt{2022LSA....11..329C,2022RAA....22j5001S}) onboard the FengYun-3E (FY-3E, \citealt{2022AdAtS..39....1Z}) satellite, with a cadence of $\sim$7 seconds and a spatial resolution of $\sim$2.5{\arcsec} pixel$^{-1}$. FY-3E, successfully launched at 23:28 UT on 4 July 2021, is the world’s first early-morning-orbit civil meteorological satellite, and X-EUVI is the first solar X-ray and EUV imager of China in orbit. We analyzed the evolution of sunspots and the magnetic field of AR 12887 by combining the intensitygrams and magnetograms observed by the Helioseismic and Magnetic Imager (HMI, \citealt{2012SoPh..275..207S}) onboard the SDO. All the SDO data used in this study were processed by the SolarSoftWare (SSW) standard procedures\footnote{\url{https://www.mssl.ucl.ac.uk/surf/sswdoc/}}. 

To investigate the quadrupolar magnetic configuration of AR 12887, the photospheric vector magnetograms were also used to extrapolate the coronal magnetic fields in this work. The photospheric vector magnetic fields were computed by the Very Fast Inversion of the Stokes Vector code \citep{2011SoPh..273..267B}, and the remaining 180{\arcdeg} azimuth ambiguity was resolved by the Minimum Energy method \citep{1994SoPh..155..235M,2009SoPh..260...83L}. We used the vector magnetograms provided by the Space-weather HMI Active Region Patches (SHARP, \citealt{2014SoPh..289.3549B}) with a cadence of 12 minutes from 00:00 UT to 19:48 UT on October 28, 2021, for extrapolations and free energy calculations of AR 12887. At first, all vector magnetograms from photospheric measurements were preprocessed to meet the force-free assumption \citep{2006SoPh..233..215W}. The nonlinear force-free magnetic field (NLFFF) with the ``weighted optimization" method of \citet{2000ApJ...540.1150W} and \citet{2004SoPh..219...87W} then was used to obtain the 3D coronal magnetic fields. The calculation was conducted within a box of $232\times208\times232$ uniform grid points with $\Delta x = \Delta y = \Delta z = 1${\arcsec}, which was downscaled from the original resolution of 0.5{\arcsec}pixel$^{-1}$. As a result, we obtained a series of extrapolations over every 12 minutes. Based on the NLFFF extrapolation of the coronal magnetic field, the distributions of the twist number $T_{w}$ \citep{2006JPhA...39.8321B} and the squashing factor $Q$ \citep{1996A&A...308..643D, 2002JGRA..107.1164T} were calculated by the code developed by \citet{2016ApJ...818..148L}. 

This work also used solar region summary and events documents\footnote{\url{ftp://ftp.swpc.noaa.gov/pub/warehouse/}} of the National Oceanic and Atmospheric Administration (NOAA) Space Weather Prediction Center to statistically analyze the area of the ARs that produced M- and X-class flares from August 1996 to November 2021. Although some omissions and inaccuracies were found in the NOAA documents, it would not affect the overall results of this study due to the large number of statistical samples. Geostationary Operational Environmental Satellites (GOES) soft X-ray (1-8 {\AA}) flux\footnote{\url{http://www.ngdc.noaa.gov/stp/satellite/goes/index.html}} is employed to determine the class of the X-ray flares.

\section{Results} \label{sec:results}
On October 28, 2021, the Sun produced a GOES X1.0 flare, the second X-class flare of Solar Cycle 25, from a small AR NOAA 12887. According to the NOAA-reported AR areas (in millionths of a solar hemisphere or $\mu h$) and flare events, the average area of AR 12887 (313 $\mu h$) is much smaller than the average area of the X-class flare-productive ARs over the past 25 years (781 $\mu h$) and even smaller than that of the M-class flare-productive ARs (504 $\mu h$). If an X-class flare occurs in a small AR, it may indicate that there is a large amount of magnetic free energy that can be released and converted into the radiation and other forms of energy. Fortunately, unlike the first X-flare of Solar Cycle 25 that exploded on the solar west limb, AR 12887 appeared near the center of the solar disk at location S26W07 on October 28 (Figure \ref{fig:fig1}(b1)-(b2)), which provided an opportunity to study the evolution of the magnetic field configuration of the AR.

\begin{table}
\renewcommand{\arraystretch}{0.88}
\centering
\caption{A flare list of AR 12887, including date, GOES flare class, time, and duration. AR 12887 erupted an X-class, 3 M-class, and 26 C-class flares during its passage through the solar disk.}
\label{tab1}
\begin{tabular}{cccccc}     
  
   \hline 
   \hline  
Date& GOES &      &Time (UT)&  & Duration \\
   \cline{3-5}
(UT)& Class & Start & Peak & End& (minutes) \\
   \hline
2021 Oct 26 & C1.9 & 00:12 & 00:19 & 00:30 & 18 \\
2021 Oct 26 & C7.8 & 05:56 & 06:04 & 06:10 & 14 \\
2021 Oct 27 & C1.0 & 03:50 &  03:57 & 04:03 & 13 \\
2021 Oct 27 & C2.8 & 05:22  & 05:35   &   05:41 & 19 \\
2021 Oct 27 &   C3.1  & 05:41  & 05:45       &05:47   & 6 \\
2021 Oct 27 &  C8.5  &05:47    &06:08      & 06:19  & 32 \\
2021 Oct 27 &  C1.5  &07:01    &07:10       &07:14  & 13 \\
2021 Oct 27 &  C2.8  &09:50    &09:54       &09:59  & 9 \\
2021 Oct 27 &    C1.1  &13:35    &13:42      &13:48  & 13 \\
2021 Oct 27 &    C1.6  &22:07    &22:22     &22:27  & 20 \\
2021 Oct 27 &   C1.1   &22:35    &22:44       &23:00  & 25 \\
2021 Oct 28 & C1.2 & 01:32 &  01:38    &  01:45  & 13 \\
2021 Oct 28 &  M1.4  &07:30  & 07:40   &   07:45  & 15 \\
2021 Oct 28 &   C1.1  &09:56  & 10:00   &   10:05  & 9 \\
2021 Oct 28 &    M2.2 &10:19  & 10:28   &   10:37  & 18 \\
2021 Oct 28 &   C1.2  &12:09  & 12:17    &  12:27  & 18 \\
2021 Oct 28 &  C3.3  &13:15   &13:21    &  13:25  & 10 \\
2021 Oct 28 &   C3.8 & 13:46 &  13:59    &  14:05  & 19 \\
2021 Oct 28 &   X1.0 & 15:17   &15:35  &    15:48  & 31 \\
2021 Oct 29 &   C1.1 &   07:26 &  07:45   &   08:05  & 39 \\
2021 Oct 29 &   C2.5 &   13:20 &  13:30   &   13:40  & 20 \\
2021 Oct 30 &  C1.0 &  21:22  & 21:29  &  21:35  & 13 \\
2021 Oct 31 &  C1.2  & 18:16   &18:27     & 18:32  & 16\\
2021 Oct 31 &  C3.2  & 19:51   &20:00    &  20:04 & 13 \\
2021 Nov 1  &   M1.5  &00:57   &01:45     & 02:10  & 73 \\
2021 Nov 1  &   C1.3 &17:28   &18:01    &  18:44  & 76 \\
2021 Nov 1  &    C4.0 &21:05   &21:33    &  21:51  & 46 \\
2021 Nov 2 &  C1.0   &  12:55   &13:05   &   13:16  & 21 \\
2021 Nov 3 &   C5.2  & 20:56  & 21:17   &   21:40  & 44 \\
2021 Nov 4 &  C1.0 & 12:25  & 12:35    &  12:41  & 16 \\
   \hline
\end{tabular}
\end{table}

\begin{figure*}[!htb]
\plotone{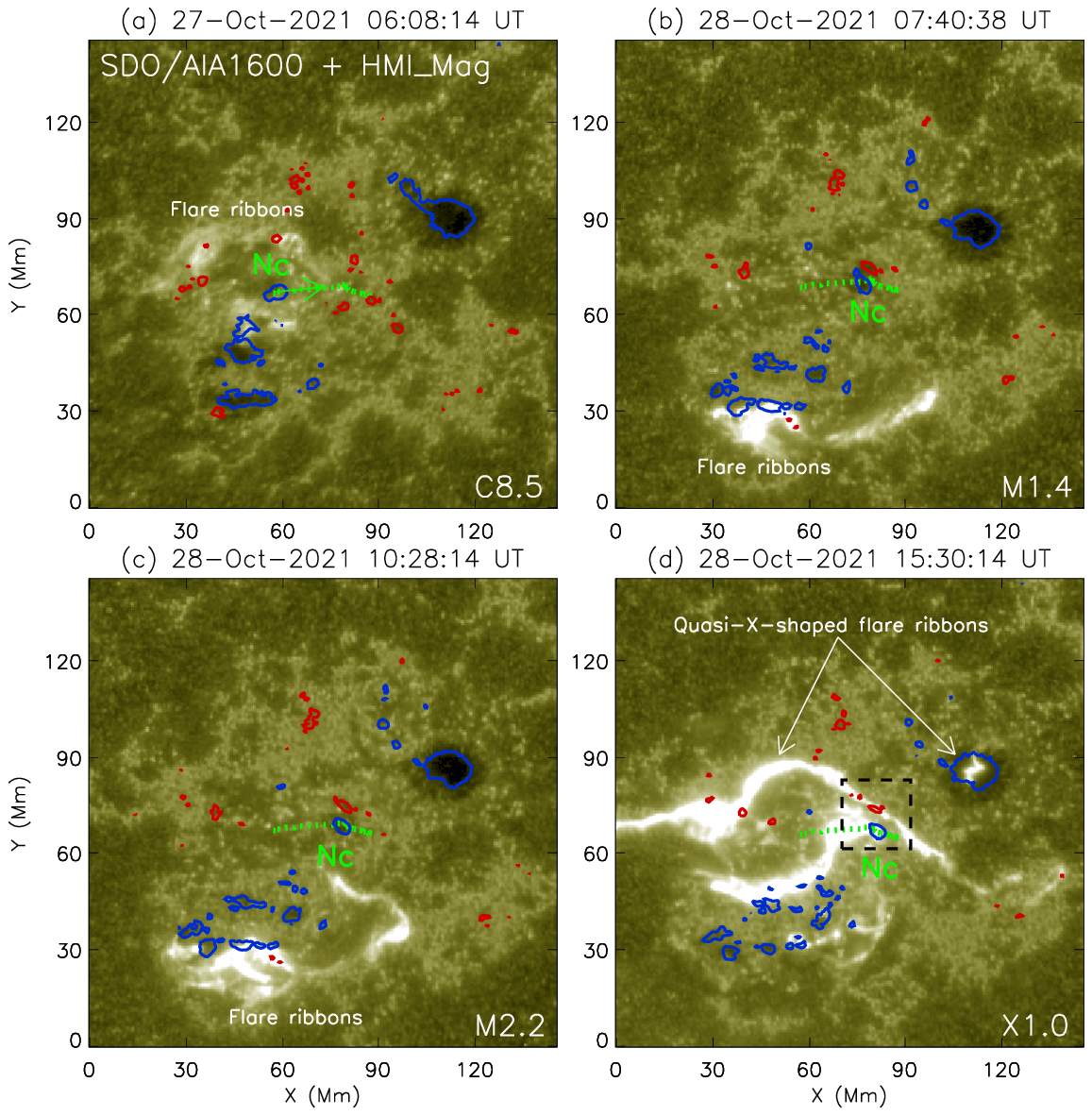}
\caption{Motion of Nc and four larger flares in AR 12887. The locations of flare ribbons are displayed by enhancing brightness on the SDO/AIA 1600 {\AA} images. The contours have the same meaning as that in Figure 1. Green dashed lines show the trajectory of Nc. The black dashed box in (d) represents the FOV of Figure \ref{fig:fig3}.
\label{fig:fig2}}
\end{figure*}

\begin{figure*}[!htb]
\includegraphics[bb=0 0 450 592,width=0.6\textwidth,clip]{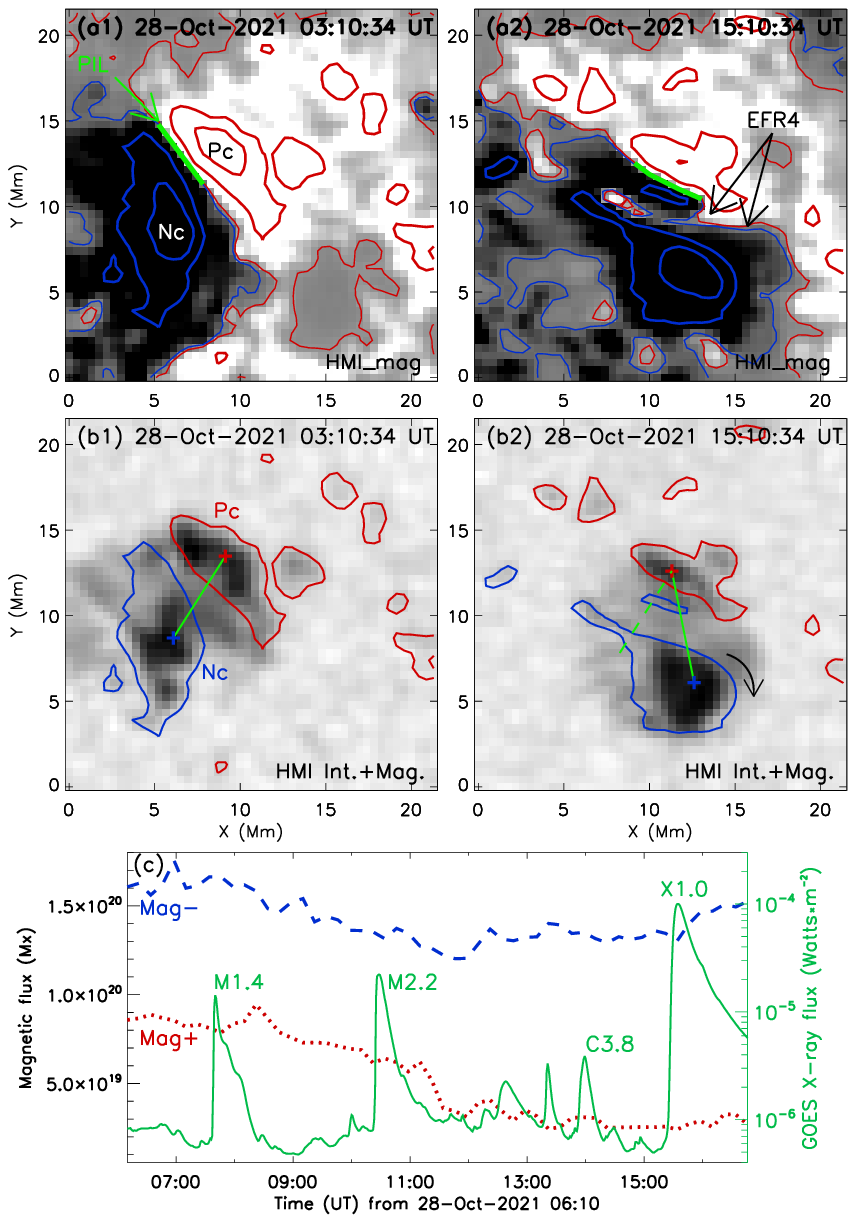}
\centering
\caption{Variation of the magnetic field around the collisional polarities (Nc and Pc). The black and white images and red and blue contours in (a1) and (a2) are SDO/HMI longitudinal magnetic fields, and the red (positive) and blue (negative) contours correspond to magnetic fields in $\pm$1000, $\pm$500, and $\pm$30 G. The thick green lines in (a1) and (a2) represent the main polarity inversion line (PIL). The background images and contours in (b1) and (b2) display HMI intensitygrams and magnetograms, respectively, and the contours correspond to $\pm$ 500 G. The blue and red plus signs mark the positions of the centers of Nc and Pc, respectively, and the green solid and dashed lines show the angle of their relative motion. The red dotted and blue dashed curves in (c) represent the positive and negative photospheric magnetic fluxes of the collisional polarities, respectively. The green curve in (c) represents the GOES soft X-ray 1-8 {\AA} flux of the Sun. A 5-second animation of panels (a1)-(b2) is available and shows the evolution of SDO/HMI white light and longitudinal magnetic fields in the polarity collision region from 23:58 UT on October 27 to 19:10 UT on October 28, 2021. We can see the magnetic emergence and sunspot rotation during the collision.
\label{fig:fig3}}
\end{figure*}

\subsection{The quadrupolar field of AR 12887} 
We analyze the evolution of the sunspots and magnetic field in AR 12887 from October 24 to 28, 2021. During its evolution, the main part of AR 12887 is a 3D quadrupolar magnetic configuration composed of two oppositely directed dipoles that are not co-linear, as shown in Figure \ref{fig:fig1}. The photospheric magnetic fields of AR 12887 are marked using red and blue contours in Figure 1(a1)-(a4), and the evolution of the AR is analyzed in conjunction with the sunspots in white light images of Figure 1(a1)-(a3).

\begin{figure*}[!htb]
\plotone{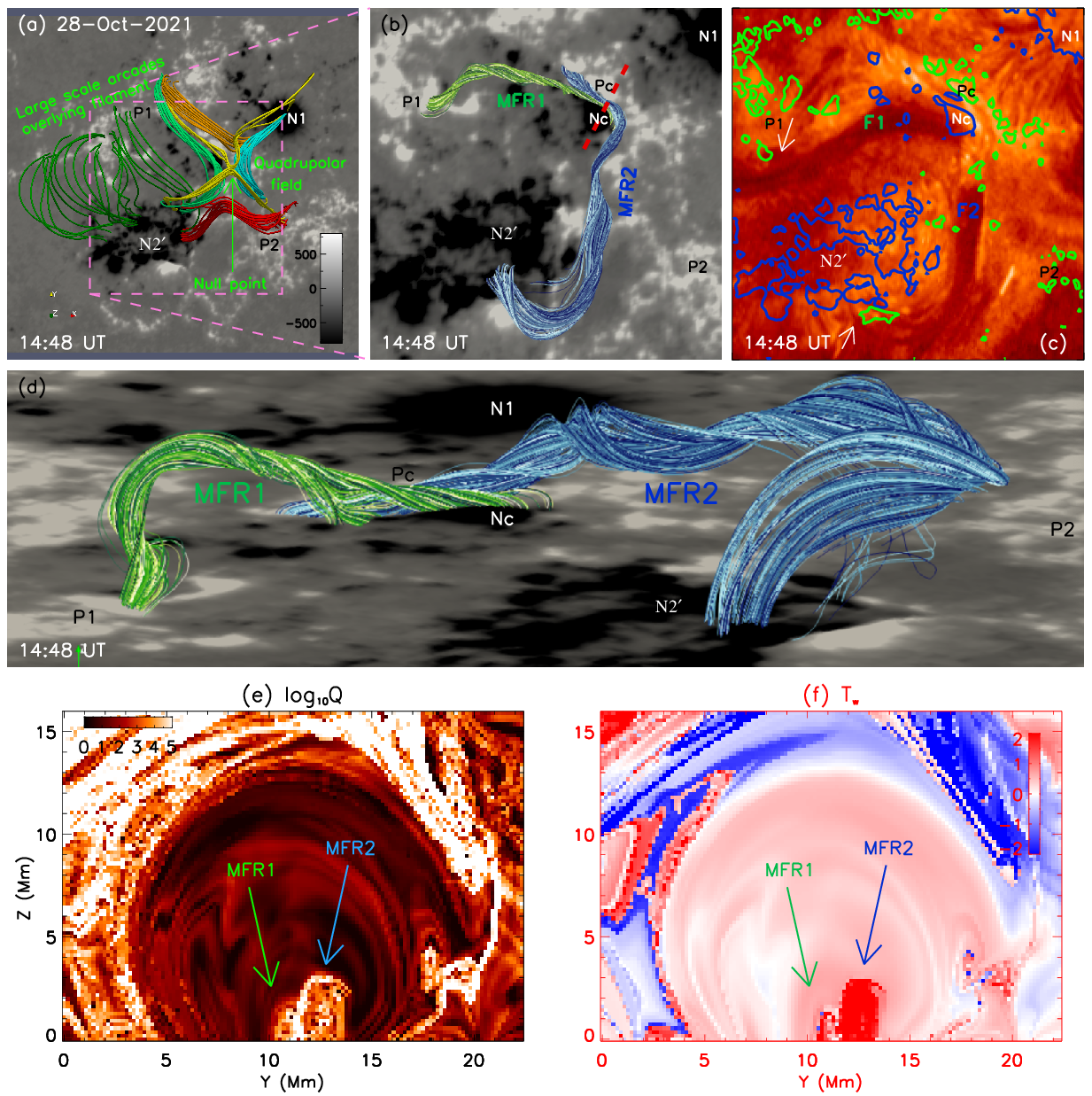}
\caption{Magnetic configuration and analysis of two interacting magnetic flux ropes (MFRs) in the extrapolated coronal magnetic fields of AR 12887 at 14:48 UT on October 28, 2021. (a) displays the large-scale arcades and 3D quadrupolar fields above the two MFRs, and the colored lines represent extrapolated coronal magnetic field lines, with different colors used to distinguish magnetic field lines at different locations of the quadrupole field and large-scale arcades overlying filament. (b) and (d) show MFR1 and MFR2 from different viewing angles. For comparison with the extrapolated fields in (b), (c) shows the distribution of two filaments (F1 and F2) in a simultaneously AIA 304 {\AA} image, and the white arrows indicate two important footpoints of the filaments. Green and blue contours indicate positive and negative magnetic fields, respectively. (e)-(f) display vertical cross sections of $Q$ and $T_{w}$ along a slice of the red dashed line shown in (b), respectively.\label{fig:fig4}}
\end{figure*}

On October 24, AR 12887 had two large negative sunspots and several small positive sunspots. Considering the distribution of the 3D quadrupolar field, we mark the magnetic structures of the AR as four poles located at different positions, as shown in Figure \ref{fig:fig1}(a1), using N1 to denote the negative magnetic structures located in the northwest part of the AR, P1 for the positive magnetic structure located in the northeast part, and N2 (P2) for the negative (positive) magnetic structures in the southeast (southwest) part of the AR. Because the magnetic structures marked here are based on the location, P1 (P2) and N1 (N2) are related, but not strictly conjugated. For example, the magnetic flux of P1 comes not only from the positive magnetic flux corresponding to N1 but also partly from the positive flux related to N2 (see also Figure \ref{fig:fig4}(a)). 

From October 25 to 26, several emerging flux regions (EFRs) appeared in the AR, while the large sunspot N2 split into Nc and other small negative sunspots (labeled as N2$^{\prime}$ in Figure \ref{fig:fig1}(a2)-(a3)). Nc may also include the negative component of the EFR2. On October 26, Nc began to move from the east to the west of the AR. We track the motion of Nc, and the blue dashed line in Figure \ref{fig:fig1}(a3) represents the trajectory of Nc from 21:58 UT on October 26 to 23:58 UT on October 28. Finally, at the center of the quadrupolar magnetic configuration, the fast-moving Nc encountered a small opposite-polarity sunspot (Pc) that may come from the positive polarity of the EFR3 (Figure \ref{fig:fig1}(a3)-(a4)). Nc and Pc come from different EFRs and these two magnetic structures with opposite polarities collide at the center of the AR, indicating the evolution of AR 12887 has entered a new stage.

Table \ref{tab1} lists the information of all flares larger than C-class in AR 12887 during its passage through the solar disk, including start, peak and end times, as well as flare classes and durations. From October 23 to November 4, a total of 30 flares erupted in the AR including an X-class, 3 M-class, and 26 C-class flares. Their duration is 6 to 76 minutes, with an average of 23 minutes. And the X1.0 flare on October 28 is the largest flare of the AR. 

From October 26 to 28, a dozen minor flares ($\textless$M1.0) erupted from the AR, but the major flares occurred only when Nc approached the core of the quadrupolar field and collided with the opposite polarity Pc. Before the X1.0 flare, the three largest flares that erupted in the AR were C8.5, M1.4, and M2.2. From the SDO/AIA 1600 {\AA} images, we can see that the bright ribbons of the C8.5 minor flare are located around Nc (Figure \ref{fig:fig2}(a)) when Nc has just started to move westward. It is noteworthy that the trajectory of Nc changed the direction of motion on October 28 and the major flares occurred. The flare ribbons of two M-class major flares were located in the southern part of AR 12887, which may correspond to the magnetic emergence and magnetic shear in this area (Figure \ref{fig:fig2}(b)-(c)). As the interaction between the Nc and Pc developed, one flare ribbon of the M2.2 flare was also closer to the center of the AR. 

Eventually, at 15:17 UT on October 28, the X1.0 flare started, and then its flare ribbons swept across the central area of AR 12887, while a remote flare ribbon crossed the middle of large sunspot N1 (Figure \ref{fig:fig2}(d)). If we consider the main flare ribbons and the remote ribbon as an integrated system, it may be regarded as a quasi-X-shaped flare ribbon. One can see that Nc and Pc, namely the collisional polarities, locates at the center of the quasi-X-shaped flare ribbon and the 3D quadrupolar field. At that time, N1 was the largest sunspot of the AR, and it had not been affected by the flare ribbons before the X1.0 flare, which implies that the fast movement and collision of Nc made N1 and the rest of the AR all engage in, and finally produced an X-class flare with a quasi-X-shaped flare ribbons.

\subsection{The magnetic fields around the collisional polarities} 
At the core of the 3D quadrupolar field, magnetic flux cancellation and relative motion occurred between Nc and Pc, and were accompanied by a rotation of Nc. The magnetic flux also emerged around the polarity inversion line (PIL) between Nc and Pc. Figure \ref{fig:fig3}(a1)-(a2) show that the morphology of the main PIL changed due to the interaction between the collisional polarities and the appearance of the magnetic emergence (for example, EFR4).  As shown in Figures \ref{fig:fig3}(b1)-(b2), during the 12 hr pre-flare period (03:10 UT-15:10 UT), Nc rotated about 44{\degr} counterclockwise relative to the center of Pc. In addition, some small magnetic structures in the penumbra of the Nc had a clockwise rotation around Nc (see the black arrow in Figure \ref{fig:fig3}(b2) and the animation of Figure \ref{fig:fig3}). 

Figure \ref{fig:fig3}(c) shows the variation of magnetic fluxes of Nc and Pc and GOES soft X-ray flux before and during the X-class flare. As shown by the blue and red curves, the negative magnetic flux of Nc was greater than the positive magnetic flux of Pc. The positive and negative magnetic fluxes both decreased until $\sim$12:00 UT on October 28. It may indicate that the positive magnetic flux was cancelled by the negative magnetic flux in this area, and the umbra area of positive sunspot Pc shrank as well as its magnetic flux. Calculations show that the average free energy of AR 12887 before the X1.0 flare (13:00 UT-14:48 UT) is about 5$\times10^{33}$ erg, which is greater than the energy of major flares. The magnetic interactions and movements of Nc and Pc indicate that the magnetic energy is accumulating in the area of the collision of polarities. 

Furthermore, we used NLFFF magnetic field extrapolation to investigate the distribution of pre-flare magnetic field in the corona at 14:48 UT on October 28, 2021. It reveals two magnetic flux ropes (MFRs) and a possible magnetic null point in the complex quadrupolar configuration of AR 12887 (see Figures \ref{fig:fig4}). From a large field of view (FOV) in Figure \ref{fig:fig4}(a1), it can be seen that AR 12887 has a magnetic configuration of the large-scale quadrupolar field with pre-existing filament magnetic systems. The top and side views of the magnetic field extrapolation results for the two MFRs in the AR are shown in Figure \ref{fig:fig4}(b) and \ref{fig:fig4}(d), respectively. 

By comparing the extrapolated magnetic field and simultaneously EUV observational images (Figure \ref{fig:fig4}(b)-(c)), we can see that one MFR (MFR1) is on the eastern part of the AR, roughly corresponding to a filament (F1) under large-scale arcades, and the other MFR (MFR2) is in the south, witch consistent with the second filament (F2). As shown by the two arrows in Figure \ref{fig:fig4}(c), we can trace the east foot point of MFR1 and the south foot point of MFR2 through the absorption of low-temperature substances in F1 and F2 in EUV images.
Note that the EUV filaments in AIA 304 \AA~seemingly were not perfect match with the corresponding MFRs, especially F1 and MFR1, which may be caused by multiple reasons. First of all, F1 is a large-scale filament, which extends beyond the eastern edge of AR 12887, and it is difficult to reproduce the twisted structure of such large-scale filament by the NLFFF method, so the large-scale arcades overlying filament can be seen from the extrapolation results, but the MFR1 is not perfect match with the corresponding F1 in AIA 304 image, especially in the eastern part of F1. Secondly, F1 is located in the southern hemisphere of the Sun, and its filament material on the EUV image may be located southerly due to the projection effect. Finally, filament mass tends to suspend only over the dipped field region of the corresponding MFR, so the extrapolated magnetic fields should not accurately reproduce the EUV images.

It is worth noting that the western part of MFR1 is rooted in Nc, while the northern part of MFR2 rooted in Pc, which means the two MFRs meet above the center of the AR. Therefore, it can be seen from Figure \ref{fig:fig4}(a)-(d) that there are two distinct MFR structures supporting two filaments in the AR. Both filaments are sinistral chirality because their axis magnetic fields are to the left when viewed from the positive polarity side of the filament channel. And their main magnetic axes are approximately perpendicular to each other. For this reason, MFR or filament description will be used interchangeably in the following text. 

We calculated magnetic twist number $T_{w}$ and squashing factor $Q$ and then analyzed their spatial distribution in cross sections. Figure \ref{fig:fig4}(e) is a vertical cross section of log $Q$ along the red dashed line shown in Figure \ref{fig:fig4}(b), and it shows that two MFRs encounter each other in this small area with a high $Q$ value (Q $\geqslant 10^{5}$) at their edges, where the current is accumulated. Figure \ref{fig:fig4}(f) shows that there are a pair of large twist number ($T_{w} \geqslant 1.75$) regions with the same $T_{w}$ sign along the slice, and they also correspond to the two MFRs. 

\begin{figure*}[!htb]
\plotone{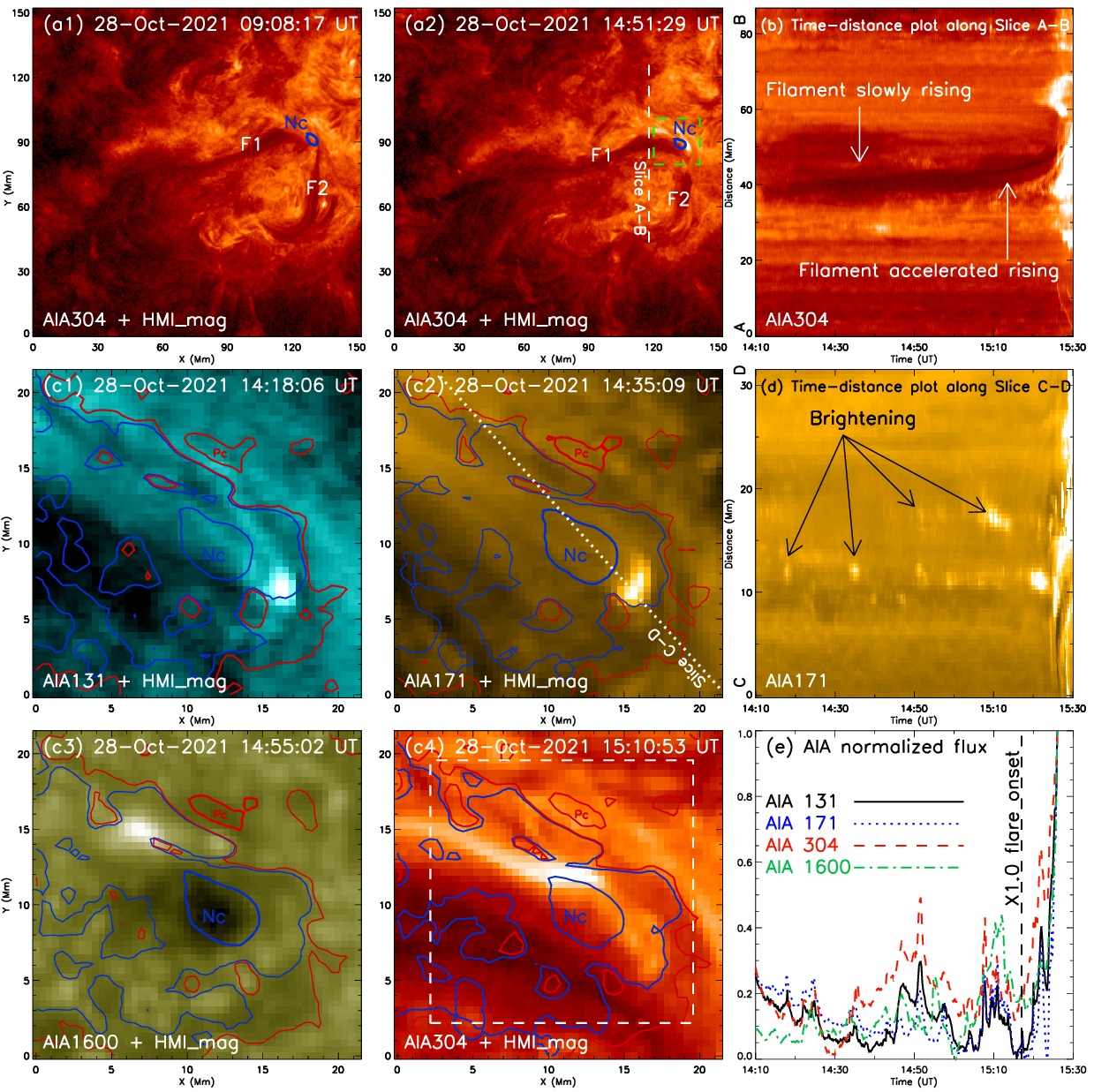}
\caption{Observations of flare triggering and erupting process with multi-passband images of SDO/AIA. (a1)-(a2) show the evolution of the two filaments (F1 and F2) before the X-class flare, and the dashed line in (a2) indicates the location of the slice shown in the time-distance diagram of (b). (c1)-(c4) display multiple brightening around Nc in AIA 131, 171, 1600, and 304 {\AA} channels, respectively. The red (30 and 800 G) and blue ($-$30 and $-$800 G) contours represent the longitudinal magnetic fields of HMI. (d) shows repeated brightening along the slice C-D in (c2). The colored lines in (e) give the normalized light curves of EUV/UV multi-passband brightness of the dashed box area in (c4).
\label{fig:fig5}}
\end{figure*}

\subsection{The onset of the X1.0 flare} 
\begin{figure*}[!htb]
\plotone{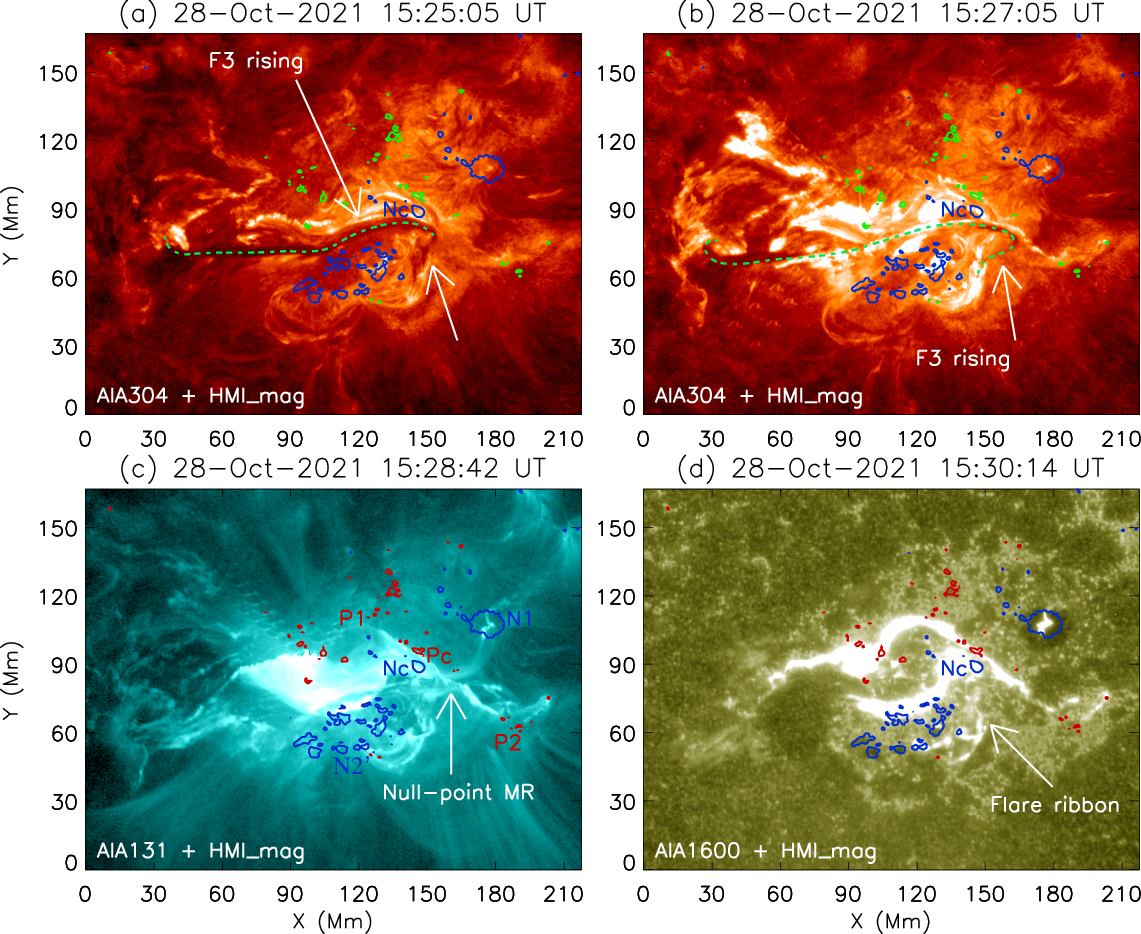}
\caption{Eruption of the newly-formed filament. (a) and (b) show the new filament (F3, see green dashed lines) rising and erupting in AIA 304 {\AA} images. In the right part of (c), the bright flaring loops may indicate the presence of null-point magnetic reconnections (MR). The flare ribbon (see the arrow) in the southern part of the AR in (d) indicates that the original F2’s material merged into F3 and participated in the eruption process. The contours represent the positive (red or green) and negative (blue) polarities at 700 and $-$700 G, respectively. A 20-second animation of panels (a)-(c) shows the evolution of SDO/AIA 304 and 131  {\AA} channel images in AR 12887 from 14:10 UT to 15:29 UT on October 28, 2021. It shows the merging process of the two filaments and the rising and erupting of the newly formed filament.
\label{fig:fig6}}
\end{figure*}

Figures \ref{fig:fig5} and \ref{fig:fig6} show the merging and eruption of the filaments, and the coronal activities before and during the X1.0 flare observed in multi-passband EUV and UV channels. With the movement of Nc, F1 on the eastern part of the AR also moves westward and gradually approaches the core of the AR, and then F1 and F2 interact at the area of the collision of polarities (Figure \ref{fig:fig5}(a1)-(a2)). 
As shown in Figures 5, 6 and the animation of Figure 6, the initial brightening occurs in the western part of the filament system, especially in the the collisional polarities region, and later expands rapidly eastward with the lifting of the filament F3.

After 14:10 UT, small-scale brightening appeared in the area around Nc. The brightening is intermittent and localized, as shown by the time-distance diagram (Figure \ref{fig:fig5}(d)) along Slice C-D in Figure \ref{fig:fig5}(c2) and normalized light curves (Figure \ref{fig:fig5}(e)) in the dashed box area of Figure \ref{fig:fig5}(c4). By a combined analysis of EUV/UV images and magnetograms (Figure \ref{fig:fig5}(c1)-(c4)), we found that the brightening occurred at the junction of these two MFRs, where opposite-polarity magnetic fluxes were colliding and canceling with each other. This interaction process results in the merging of the two distinct MFRs, which is highly suggestive of a tether-cutting-like reconnection process \citep{2001ApJ...552..833M, 2014ApJ...797L..15C, 2018ApJ...869...78C, 2019ApJ...887..118C}, and as a result, a longer filament (F3 or MFR3) gradually formed (see Figure \ref{fig:fig6}(a)-(b)). 

As shown by the time-distance diagram (Figure \ref{fig:fig5}(b)) along Slice A-B in Figure \ref{fig:fig5}(a2), it may lead to the loss-of-equilibrium of the large newly-formed filament (F3), i.e., a slow ascending and then an accelerated rising of the large filament, and eventually trigger the X-class flare. As shown in Figure \ref{fig:fig6}(a)-(b), in the process of filament rising, it can be seen that the filament material comes from the original locations of F1 and F2, also indicating that the erupting filament is a new filament (F3) formed by F1 and F2. Meanwhile, the enhancement of SDO/AIA 1600 {\AA} image shows the distribution of the flare ribbons, and it can be seen that a shorter flare ribbon appeared at the position of F2 (see the arrow in Figure \ref{fig:fig6}(d)), which also confirmed that F2 participated in the merging process, and finally led to the eruption of F3. 

As the new filament rises, two-ribbon flaring loops appear in the eastern part of the AR, and some overlying loop systems may be pushed to the western part of the AR (Figure \ref{fig:fig6}(b)-(d)). This may result in magnetic reconnection at the null point of the quadrupolar field (Figure \ref{fig:fig6}(c)) before the peak time (15:35 UT) of the X-class flare. The observed multiple brightening and the morphology of flaring loops confirm the reliability of the extrapolated coronal magnetic fields. It suggests that the standard flare model and the theory of 3D reconnection at the core of quadrupolar field are both needed to understand the X-class flare.

\section{Discussion and Conclusions} \label{sec:dc}

\begin{figure*}[!htb]
\plotone{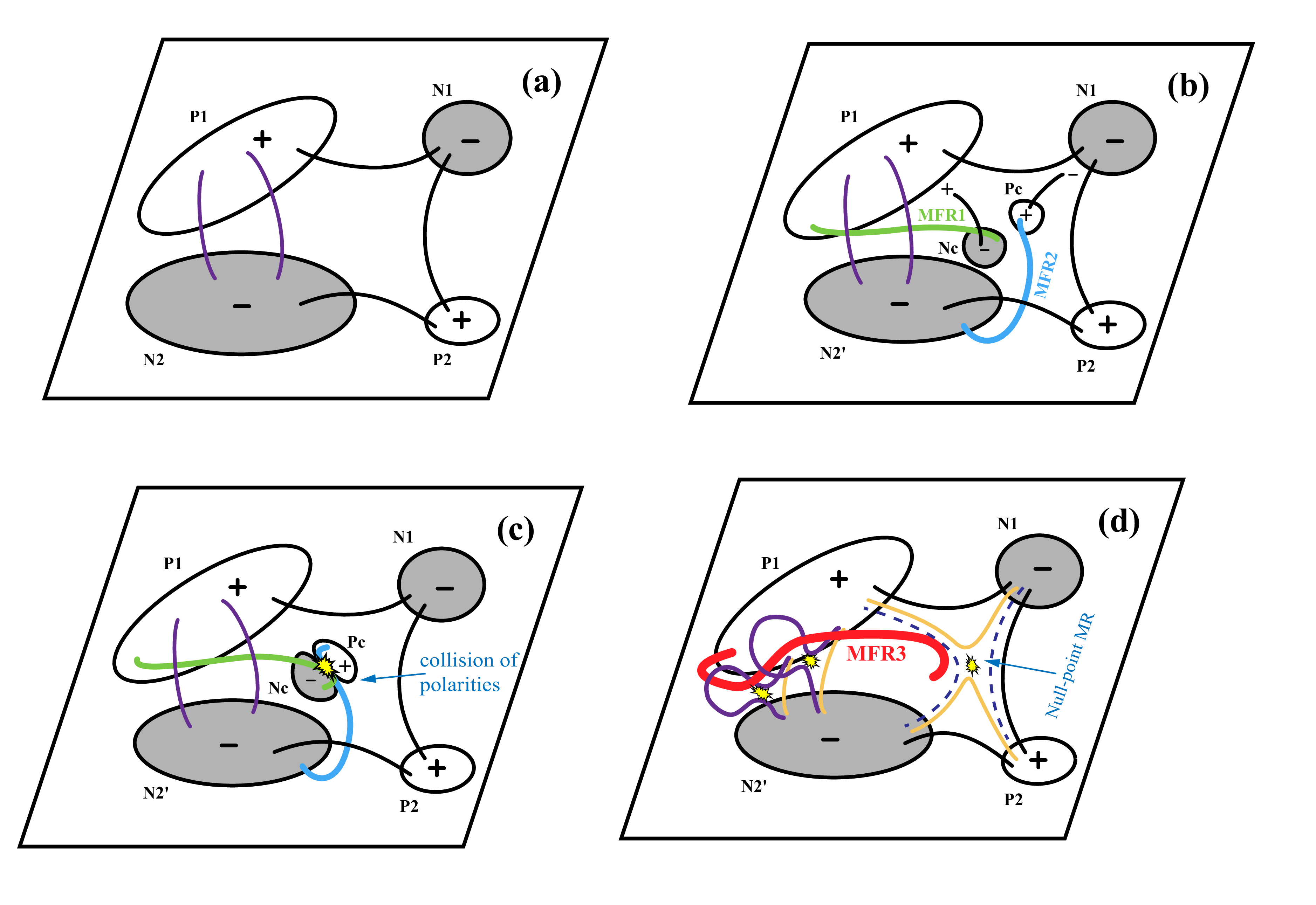}
\caption{Schematic illustrations of the magnetic field evolution and the trigger process for the X1.0 flare in AR 12887. The green and light blue lines represent two distinct MFRs (MFR1 and MFR2) before the merging process, respectively, and the red line represents the newly-formed MFR (MFR3) during the eruption. Other lines display the magnetic fields of the AR. (c) and (d) show that the magnetic reconnections (MR) occur first around the collisional polarities (Nc and Pc) where two MFRs meet and then at the null point in higher corona during the flare.
\label{fig:fig7}}
\end{figure*}

The above results converge to a clear and concise scenario shown in Figure \ref{fig:fig7}. In the beginning, the AR has a magnetic field configuration composed of a quadrupolar field and large-scale arcades overlying filament (Figure \ref{fig:fig7}(a)). Several EFRs appeared in the AR, while a large sunspot (N2) disintegrated. The magnetic configuration of a 3D quadrupolar field may be favorable for the formation of multiple filaments, but an appropriate mechanism is still needed to trigger a major eruption. One small sunspot (Nc) and one end of a filament (MFR1) fast move toward the center of the quadrupolar field (Figure \ref{fig:fig7}(b)). The sunspot Nc collides with a nonconjugated opposite magnetic polarity Pc (i.e., the collision of polarities), and another filament (MFR2) meets MFR1 at the same area. Then the tether-cutting-like reconnection occurs above the area, resulting in the merging of these two MFRs, and the rising of the newly-formed large filament (MFR3) finally triggers an energetic X-class flare (Figure \ref{fig:fig7}(c)). A two-ribbon flaring loop first appears below the rising large filament, and then the rising filament pushes the overlying magnetic fields westward and forms some flaring loops around the core of the AR (i.e., the null point) during the impulsive phase of the X-class flare. Similar scenario has also been reported in \citealt{2020A&A...640A.101H}, where an erupting filament pushed its overlying fields toward the fields above another stable filament around and caused successive external reconnection between their overlying fields, resulting in simultaneous brightenings on both sides of the stable filament. Eventually, the flaring loops of the X1.0 flare in AR 12887 display a mixed morphology of the two-ribbon flare and the X-shaped flare (Figure \ref{fig:fig7}(d)), and accompanied by a halo CME, a major solar energetic particle event, and a global EUV wave event \citep{Hou2022}. 

Quadrupolar-field ARs are found to a desire birth place of solar flares (e.g.,  AR 11967 and AR 12887), but appropriate conditions are need for a major flare \citep{2017ApJ...835..156N}. 
Previous studies tended to attribute the occurrence of major flares in quadrupolar ARs to magnetic reconnection at a null point or QSLs. In quadrupolar ARs, magnetic reconnection in these unique magnetic topologies theoretically can lead to more complex flare ribbons than bipolar ARs. Indeed, several previous studies have reported the occurrence of X-shaped major flares in quadrupolar ARs, such as NOAA 11967 \citep{2016ApJ...823L..13L, 2017ApJ...850....8J, 2017ApJ...842..106K}. 
In our current study, the X-class flare in the quadrupolar AR 12887 is found to associate with quasi-X-shaped flare ribbons, and it displayed three distinct observational features: (1) the NOAA-reported area of AR 12887 (250 $\mu h$ on the day of the X-class flare) is much smaller than that of AR 11967 (1410 $\mu h$ on the day of three X-shaped M-class flares), but AR 12887 unexpectedly erupted an X-class flare and AR 11967 only erupted M-class flares. 
(2) In our case, the flare ribbons swept the whole AR, but those flare ribbons reported in other ARs only located at the outskirts or a limited part of their corresponding AR. 
(3) The X-class flare in our case indeed included an ensuing null-point reconnection, but it appeared to be a byproduct at the overlying null point due to the merging and ensuing eruption of two MFRs. 
Based on these features, we suggest that the collision of Nc and Pc facilitated the interaction and merging of two adjacent MFRs, triggering an eruption in the quadrupolar magnetic configuration. As a result, due to the merging eruption process, additional reconnection was induced at the overlying null-point topology. Therefore, the entire AR 12887 participated in the eruption and released more amount of magnetic free energy, leading to the unexpected X-class flare.

Moreover, the collisional polarities (Nc and Pc) in AR 12887 come from different EFRs (i.e., nonconjugated polarities), which is consistent with AR 11158 and AR 12017 in a previous study \citep{2019ApJ...871...67C}. They have collisional shearing at the center of the quadrupolar field in AR 12887, accompanied by the magnetic cancellation, magnetic emergence, and sunspot rotation. The collisional polarities of AR 12887 is somewhat similar but not strictly a $\delta$ sunspot because the umbrae of opposite polarities are only partially surrounded by the penumbra. It is also small ($\sim$10 Mm) with a short main PIL. Compared with large $\delta$ spots in other ARs \citep{2005ApJ...622..722L, 2017ApJ...849L..21Y} which may store more magnetic free energy, the main function of the collisional polarities in AR 12887 may be to act as a trigger of the flare process. It has indeed played a similar role to large $\delta$ spots, providing a common PIL for the two filaments (F1 and F2), so that the magnetic axes of the originally approximately perpendicular filaments are parallel in the PIL, and eventually lead to the merging of two filaments with the same chirality. In addition, the three X-shaped flares of AR 11967 were all confined (i.e., without CME), but the X1.0 flare of AR 12887 was accompanied by a CME due to the existence and eruption of the filaments in its 3D quadrupolar configuration. 

Traditional flare forecasting methods usually use the area and the magnetic classification of an AR to quickly estimate the probability of the occurrence of a major flare. However, AR 12887 is not only small in area but also has a relatively simple magnetic classification ($\beta\gamma$ on October 28, 2021). It may lead to underprediction and confirms the importance of magnetic parameter calculations in flare forecasting.

This study focused on the X1.0 flare that occurs in the complicated magnetic configuration of AR 12887, which is a large-scale quadrupolar field accompanied by the pre-existing filament magnetic system. We found that: 1) although the area of AR 12887 is smaller than the average area of the major flare-productive ARs over the past 25 years and also smaller than those of previously reported X-shaped flare ARs, the magnetic emergence, magnetic shear, and other magnetic field evolution processes in the AR accumulate enough magnetic free energy to produce an X-class flare; 2) the main function of the collision of polarities is triggering the coupled eruption of the magnetic configuration with the pre-existing filament magnetic system and the quadrupolar field, i.e., the fast convergence and collision of nonconjugated polarities lead to two filaments (MFRs) interacting and subsequent merging into a larger filament due to the tether-cutting-like reconnection process with their opposite-polarity footpoints located on the different polarities, and then the newly-formed large filament loses its equilibrium, rises and erupts, involving the entire AR, and finally produce the mighty eruption of the X-class flare with a quasi-X-shaped flare ribbon and a CME. Therefore, this work provides a unique observational ingredient for understanding major flares in multipolar ARs and also has a practical significance for the forecasting operation of space weather.

\acknowledgments
We sincerely thank Prof. Hui Tian and Prof. Pengfei Chen for their kind and valuable suggestions. The SDO data have been used by courtesy of AIA and HMI science teams. We thank the in-orbit test and science teams of the National Satellite Meteorological Center for the FY-3E/X-EUVI data. This work was supported by the National Key R\&D Program of China (2021YFA0718600) and the National Natural Science Foundations of China (42274217, 41931073, 41774195), Ten-thousand Talents Program of Jing-Song Wang, and Yunnan Key Laboratory of Solar Physics and Space Science under the number YNSPCC202221. S.H.Y. was supported by the National Key R\&D Program of China (2019YFA0405000) and the National Natural Science Foundations of China (11790304). H.C.C. was supported by the National Nature Science Foundations of China (12103005) and Yunnan Key Laboratory of Solar Physics and Space Science under the number YNSPCC202210. Z.Y.H. was supported by the China Postdoctoral Science Foundation (2021M700246). Y.J.H. was supported by the National Natural Science Foundation of China (12273060) and the Youth Innovation Promotion Association of CAS (2023063).

\bibliography{AR2887r2}{}
\bibliographystyle{aasjournal}

\end{document}